\documentclass[a4paper,11pt]{article}
\usepackage{pos}

\usepackage{braket}

\def\cbar{\overline{c}}

\def\lbar{\overline{\ell}}

\newcommand{\dx}{\mathrm{d}}

\title{Heavy sterile neutrinos in $B$ decays and new QCD corrections to their semi-hadronic decay rates}
\ShortTitle{HSN in $B$ decays and new semi-hadronic QCD corrections}

\author*{Tim Kretz}

\affiliation{Institute for Theoretical Particle Physics (TTP), Karlsruher Institute of Technology (KIT)\\
  Wolfgang-Gaede-Straße 1, 76131 Karlsruhe, Germany}

\emailAdd{tim.kretz@kit.edu}

\abstract{In modern experiments on flavour physics it is possible to search for the decays of $B$ or $D$ mesons or $\tau$ leptons into final states with heavy neutrinos $N$ (a.k.a. heavy neutral leptons). I present a common study of theorists and experimentalists from Belle II on constraints on $B \rightarrow D^{*} \ell N$. Next I discuss the status of the theory predictions of the various $N$ decay rates. In scenarios in which $N$ interacts with SM particles only through sterile-active neutrino mixing, the dependence of
the lifetime on the relevant mixing angles is important to determine whether $N$ decays in the detector or outside. To calculate the inclusive decay rate into semi-hadronic final states reliably one needs to include radiative QCD corrections. I present analytic results for the QCD-corrected  decay rates and discuss their phenomenological impact.   
}

\FullConference{The European Physical Society Conference on High Energy Physics (EPS-HEP2025)\\
7-11 July 2025\\
Marseille, France\\}

\begin{document}
\maketitle

\section{Overview}
Heavy sterile neutrinos (HSN), sometimes also referred to as heavy neutral leptons (HNL), are interesting to study, because of their role in theories of neutrino masses \cite{Yanagida:1979as,Minkowski:1977sc}, leptogenesis \cite{Fukugita:1986hr,Davidson:2002qv}, or Dark Matter \cite{Asaka_2005,Asaka_2005_2}. 

In this work I discuss HSN production in the decay $B \rightarrow D^{*} \ell N$ and its impact on recent measurements of angular coefficients made by Belle II \cite{Belle:2023bwv}. Furthermore I will discuss inclusive hadronic decay modes of HSN. Calculations of exclusive multi-hadron final states are currently out of reach. We show how to calculate the inclusive decay rate of HSN decaying into a lepton and a hadronic system $N \rightarrow \ell + \mathrm{had.}$ utilizing known results of electroweak gauge boson correlators up to the five-loop level of QCD.

The most common way in which HSN are usually introduced is via a mixing angle (see. \cite{king2025righthandedneutrinosseesawmodels} for an overview)
\begin{equation}
\nu_L^\ell = \nu \cos\theta  + N^c \sin\theta
\end{equation}
which arises naturally in e.g. see-saw type I models by adding a right-handed neutrino field to the Standard Model (SM). Here $\theta$ is the mixing angle and $\nu_L^\ell$ denotes the left-handed neutrino field in the interaction basis with $\ell$-lepton flavor, while $\nu$ and $N$ denote the mass basis. This is the model we employ for the calculation of the inclusive HSN QCD corrections. Although the mixing angle is the most common approach to HSN this does not exhaust all possible interactions with known particles. The most general description of quarks interacting with HSN involves dimension-6 operators \cite{Robinson_2019}. We use them for the calculation of the angular coefficients of the decay $B \rightarrow D^{*} \ell N$.

\section{Sterile neutrinos from $B \rightarrow D^{*} \ell N$}
In this section I discuss the results of Ref.~\cite{Bernlochner:2024xiz}

The semi-leptonic decay $B \rightarrow D^{*} \ell \nu$ contains a neutrino in the final state which cannot be detected directly. Any measurements of this decay could thus contain a contamination of a decay involving a HSN of the type $B \rightarrow D^{*} \ell N$. Since the SM neutrino is not detected directly the effects of a heavier sterile neutrino would only be visible indirectly in precise measurements of the angular distributions of the decay. To calculate the effect of the HSN we employ the following dimension-6 operators \cite{Robinson_2019}
\begin{align}
\mathcal{H}_{\rm eff} = \frac{4G_F}{\sqrt2} V_{cb} & \left[ (\cbar_L \gamma_\mu b_L) (\lbar_L \gamma^\mu \nu_{\ell,L}) + 
g_{V_R}^{N,\ell} (\cbar_R \gamma_\mu b_R) (\lbar_R \gamma^\mu N_R)
+ g_{S_L}^{N,\ell} (\cbar_R b_L) (\lbar_L N_R) \right. \nonumber \\
& + g_{S_R}^{N,\ell} (\cbar_L b_R) (\lbar_L N_R)
  + \left. g_{T}^{N,\ell} (\cbar_L \sigma_{\mu\nu} b_R) (\lbar_L \sigma^{\mu\nu} N_R) \right] + \mathrm{h.c.}\, ,
\label{eq:hamiltonian}
\end{align}
where the $g_X^{N, \ell}$ denote the Wilson Coefficients (WC) associated with the operator $X$ and the lepton flavor $\ell$. A mixing angle in the language of operators would correspond to a dimension-7 operator and is thus neglected from our analysis. Using these operators we adapt the calculation in Ref.~\cite{Gratrex:2015hna} to calculate the fully differential decay rate
\begin{alignat}{1}
\label{eq:d4GJi}
\frac{32 \pi}{9} \frac{d^4 \Gamma}{dq^2\,d\textrm{cos}\theta_{\ell} \, d\textrm{cos}\theta_V \, d \phi }  = & \left(J_{1s} + J_{2s}\cos 2\theta_{\ell} + J_{6s} \cos \theta_{\ell} \right) \sin^2 \theta_V \,+  \nonumber \\
&{} \left(J_{1c} + J_{2c}\cos 2\theta_{\ell} + J_{6c} \cos \theta_{\ell} \right) \cos^2 \theta_V \,+  \nonumber \\
&{} \left(J_3 \cos 2 \phi +J_9  \sin 2\phi\right)\sin^2 \theta_V \sin^2 \theta_{\ell} \, +  \nonumber \\
&{} \left(J_4 \cos  \phi +J_8  \sin \phi\right) \sin 2 \theta_V \sin 2\theta_{\ell}  \,+  \nonumber \\
&{} \left(J_5 \cos  \phi +J_7  \sin \phi\right) \sin 2 \theta_V \sin \theta_{\ell} \, , 
\end{alignat} 
where the $J_i$ are the angular coefficients. Since both the decays $B \rightarrow D^{*} \ell \nu$ and $B \rightarrow D^{*} \ell N$ have distinct final states 
the angular coefficient separates into an SM and a New Physics (NP) part
\begin{equation}
J_i = J_i^{\mathrm{SM}} + J_i^{\mathrm{NP}}(g_X^{N,\ell}).
\end{equation}
The angular coefficients have been measured by the Belle II experiment and the data is available separately for $\ell = e$ and $\ell = \mu$ \cite{Belle:2023bwv} allowing us to perform the analysis for both lepton flavors independently. An important detail to consider are the experimental limitations on the probable HSN mass range. The Belle II analysis of the angular coefficients was performed in such a way that HSN effects would only be visible in a mass range up to roughly $m_N = 62.5 \, \mathrm{MeV}$.
We performed a Bayesian analysis fitting our angular coefficients to the Belle II data utilizing the \texttt{HEPfit} code \cite{de_Blas_2020}. We varied the WCs, the HSN mass $m_N$, and the form factors (FF). The fits were performed in two ways: First, setting all WCs to zero except for one and, second, keeping all WCs non-zero. We assumed flat priors for all fitted parameters. In both cases the mass posterior was completely flat over the entire mass range $m_N \in [0,62.5] \, \mathrm{MeV}$. Furthermore the fits show no preference for any specific FF parametrization (our results are reported for the JLQCD FF \cite{Aoki:2023qpa} but we checked whether using the HPQCD FF \cite{Harrison:2023dzh} or FNAL/MILC FF \cite{FermilabLattice:2021cdg} as priors leads to significantly different FF posteriors. It turns out that the FF posterior is largely insensitive to the FF prior choice). There is full agreement with the SM for all WCs at the $\sim 2 \sigma$ level i.e. all highest posterior density intervals agree with a zero WC at the $95.45 \, \%$ level. HSN masses $m_N > 62.5 \, \mathrm{MeV}$ may rudimentarily be probed using the missing mass squared distribution in Ref.~\cite{Belle:2023bwv}. We performed a bump hunt on the digitized spectrum with the template of a HSN in the shape of the peak of a SM neutrino scanning over the mass range. We find a locally peaked $p$-value at a mass of $m_N = 354 \, \mathrm{MeV}$. This is an interesting finding though not statistically significant.

\section{Hadronic sterile neutrino decay}
In this section I discuss the results of Ref. \cite{kretz:wpaper}, in which also HSN in $\tau$ decays are covered.  

Employing the mixing angle description of HSN it is possible for sterile neutrinos to decay similarly to $\tau$-leptons i.e. via the $W$ boson. Analogously to $\tau$-decay this means that for a significantly heavy sterile neutrino the decay $N \rightarrow \ell + \mathrm{had.}$ is possible through the virtual decay $W \rightarrow u' \bar{d}'$ where $u'$ and $d'$ describe any up- or down-type quark permitted by the kinematics. The decay $W \rightarrow u' \bar{d}'$ correctly describes the inclusive decay $W \rightarrow \mathrm{had.}$ provided that enough corrections in orders of the strong coupling constant $\alpha_s$ are included. Virtual decays of gauge bosons are fully described by the correlator \cite{BRAATEN1992581,Chetyrkin_1998}
\begin{align}
  \Pi^{V/A}_{\mu\nu, \, ij}(q,m_i,m_j,\mu,\alpha_s)
  &= i \int \dx x e^{iqx} \braket{0|
     \hat{T}\{ j^{V/A}_{\mu, \, ij}(x) j^{V/A \, \dagger}_{\nu, \, ij}(0)  \}|0} \\
  &= (-g_{\mu\nu} q^2 + q_\mu q_\nu) \Pi^{(1)}_{ij, \, V/A}(q^2)
    + q_\mu q_\nu \Pi^{(0)}_{ij, \, V/A}(q^2), \label{eq:pva}
\end{align}
which has been calculated, including QCD corrections, up to the five-loop level \cite{PhysRevLett.108.222003,Baikov:2008jh}. Here $q$ is the momentum of the gauge boson and $m_{i,j}$ are the quark masses associated to the quark fields in the currents $j^{V/A}_{\mu, \, ij} = \bar{q}_i \gamma_{\mu} (\gamma_5)q_j$. We follow the calculation in Ref.~\cite{BRAATEN1992581,Beneke_2008} and use $\Pi^{(1+0)}_{ij, \, V/A}(q^2)\equiv \Pi^{(1)}_{ij, \, V/A}(q^2) +\Pi^{(0)}_{ij, \, V/A}(q^2)$. The total inclusive decay rate reads
\begin{align}
\label{eq:incldecaywidth}
\Gamma(N \rightarrow \ell + \mathrm{had.} ) =& N_c \frac{G_F^2 m_N^5 |V_{N\ell}|^2 }{192 \pi^3} \times 12 \pi \int\limits_0^{(1-x_\ell)^2} d x \, (1+x_\ell^2 -x) \sqrt{\lambda(1,x,x_\ell^2)} \\
& \times \bigg[ \bigg(1+2x+x_\ell^2 - \frac{4 x_\ell^2}{1+x_\ell^2 -x}\bigg) \mathrm{Im}\, \Pi^{(1+0)}(m_N^2 x) - 2x \mathrm{Im}\, \Pi^{(0)}(m_N^2 x)  \bigg], \nonumber
\end{align}
here $V_{N\ell}$ denotes the mixing angle, $N_c = 3$ is the number of colors, $x=q^2/m_N^2$, $x_\ell = m_\ell/m_N$, and $\lambda(a,b,c) = a^2+b^2+c^2-2(ab+ac+bc)$ is the Källén-function. The longitudinal part of the correlator is proportional to the quark masses $\Pi^{(0)}(q^2) \propto m_i \mp m_j$ which is a consequence of a Ward identity connecting the longitudinal part of the correlator to the scalar or pseudo-scalar correlator and the quark condensate \cite{Becchi:1980vz,Pich_2021}. For all light quarks this means $\Pi^{(0)}(q^2) = 0$. Heavier quarks like $c$ or $b$ would require additional terms in an expansion in terms of $m_{c,b}^2/q^2$. We neglect this here and proceed in the chiral limit of $m_i=0$. The inclusive decay rate then reads
\begin{align}
\Gamma(N \rightarrow \ell + \mathrm{had.} ) =& N_c \frac{G_F^2 m_N^5 |V_{N\ell}|^2 }{192 \pi^3} \\
&\hspace{-2cm}\times 12 \pi \int\limits_0^{(1-x_\ell)^2} d x \, \bigg((1+x_\ell^2 -x)(1+2x+x_\ell^2) - 4 x_\ell^2\bigg) \sqrt{\lambda(1,x,x_\ell^2)} \, \mathrm{Im} \, \Pi^{(1+0)}(m_N^2 x). \nonumber
\end{align}
The correlator can be written as an expansion in $\alpha_s$ and logarithms of the gauge boson momentum \cite{Beneke_2008}
\begin{equation}
\mathrm{Im}\, \Pi^{(1+0)}(m_N^2 x) \propto \sum\limits_{n=0}^{\infty} a_{m_N}^n \sum\limits_{k=0}^{n+1} c_{n,k} \, \mathrm{Im} \, \ln^k(-x), \label{eq:imp}
\end{equation}
where $a_\mu = \alpha_s(\mu)/\pi$ and $c_{n,k}$ are numbers \cite{Chetyrkin:1979bj,PhysRevLett.43.668,PhysRevLett.44.560,Gorishnii:1988bc,Kataev:1990kv,Surguladze:1990tg,Gorishnii:1990vf,Baikov:2008jh,PhysRevLett.108.222003,BRAATEN1992581}. This leads to integrals of the form
\begin{equation}
I_k = \int\limits_0^{(1-x_\ell)^2} d x \, \bigg((1+x_\ell^2 -x)  (1+2x+x_\ell^2) - 4 x_\ell^2 \bigg) \sqrt{\lambda(1,x,x_\ell^2)} \ln^k( x). \label{eq:defik}
\end{equation}
We calculated this integral and found closed form results in $x_\ell$ involving dilogarithms and trilogarithms up to $k\leq 2$. For $k\geq 3 $ we use a new series representation of the Källén-function, see Ref. \cite{kretz:wpaper}. The full semi-hadronic width then becomes
\begin{align}
\Gamma(N \rightarrow \ell + \mathrm{had.}) =& N_c \frac{G_F^2 m_N^5 |V_{N\ell}|^2 }{192 \pi^3} \nonumber\\
& \times 2 (|V_{ud}|^2 + |V_{us}|^2) \bigg[ I_0 c_{0,1} + a_{m_N}  c_{1,1} I_0 \nonumber \\
&+a_{m_N}^2 \big( c_{2,1} I_0 +2 c_{2,2} I_1 \big) \nonumber\\
&+a_{m_N}^3 \big(  c_{3,1} I_0 +2 c_{3,2} I_1 - (\pi^2 I_0-3  I_2 )c_{3,3} \big) \nonumber\\
&+a_{m_N}^4 \big(  c_{4,1} I_0+2 c_{4,2} I_1 \nonumber\\
&\hphantom{a_\mu^4 }
 -(\pi^2 I_0-3 I_2)c_{4,3} - (4\pi^2 I_1 -4 I_3 )c_{4,4} \big) 
\bigg].
\end{align}
We use this result to analyze the range of HSN masses $m_N$ leading to a stable perturbative expansion. We therefore require that higher orders in $\alpha_s$ yield smaller contributions and decreased sensitivity on the renormalization scale $\mu$. We find that for $\ell=e$ and $\ell=\mu$ the perturbative expansion is stable, i.e. insensitive to the renormalization scale, for HSN masses $m_N \gtrsim 1.5 \, \mathrm{GeV}$. For $\ell = \tau$ we find a stable perturbative description for $m_N \gtrsim 3 \, \mathrm{GeV}$. In the indicated mass ranges we can therefore predict the $W$-mediated contribution to the total width reliably, which is a prerequisite for the calculation of branching ratios.


\section{Summary}
The measurement of the decay $B \rightarrow D^{*} \ell \nu$ could be diluted by the additional decay into HSN $B \rightarrow D^{*} \ell N$. The signatures for both decays would be the same except for subtle differences in the kinematics, which would be visible in the angular distributions. We calculated the differential decay width for the decay $B \rightarrow D^{*} \ell N$ using dimension-6 operators and fitted it to the recent Belle II data on the angular distributions. We performed our analysis separately for $\ell = e$ and $\ell =\mu$. We find no evidence for NP in the angular distributions in a mass range of $m_N \in [0,62.5] \, \mathrm{MeV}$. Outside of this mass range the angular distribution data is not usable, as it is biased towards the SM. In addition we performed a bump hunt on the digitized missing mass squared distribution utilizing a SM neutrino template and find the most significant local $p$-value at a mass of $m_N = 354 \, \mathrm{MeV}$. This is however not statistically significant and more thorough work and a full analysis would be required. 

We calculated the total decay rate of $N \rightarrow \ell + \mathrm{had.}$ in the mixing angle scenario. To that end we used the known results about gauge boson correlators to calculate the decay rate up to $\alpha_s^4$. We find closed form results in terms of polylogarithms up to $\alpha_s^3$ and at $\alpha_s^4$ we use a new series representation of the Källén-function. We use these results to determine the mass range of HSN masses for which the perturbative expansion is stable i.e. by requiring that higher order corrections in $\alpha_s$ yield smaller corrections and that the sensitivity on the renormalization scale decreases. We find that perturbation theory is applicable for masses $m_N \gtrsim 1.5 \, \mathrm{GeV}$ if the produced lepton is either $e$ or $\mu$ or $m_N \gtrsim 3 \, \mathrm{GeV}$ if the produced lepton is a $\tau$.

\newpage
\acknowledgments
I thank the organizers of \emph{EPS-HEP2025} for the opportunity to give this parallel talk. I am grateful for the fruitful and enjoyable collaboration with Florian U. Bernlochner, Marco Fedele, Ulrich Nierste, and Markus T. Prim on the work presented in Ref.~\cite{Bernlochner:2024xiz,kretz:wpaper}. I thank Ulrich Nierste for his support. This research was supported by the Deutsche Forschungsgemeinschaft (DFG, German Research Foundation) under grant 396021762 - TRR 257 for the Collaborative Research Center \emph{Particle Physics Phenomenology after the Higgs Discovery (P3H)}.


\bibliographystyle{JHEP}
\bibliography{proceedingsEPS}

@article{Beneke_2008,
   title={$\alpha_s$ and the $\tau$ hadronic width: fixed-order, contour- improved and higher-order perturbation theory},
   volume={2008},
   ISSN={1029-8479},
   url={http://dx.doi.org/10.1088/1126-6708/2008/09/044},
   DOI={10.1088/1126-6708/2008/09/044},
   number={09},
   journal={Journal of High Energy Physics},
   publisher={Springer Science and Business Media LLC},
   author={Beneke, Martin and Jamin, Matthias},
   year={2008},
   month=sep, pages={044–044} }

@article{Chetyrkin_1998,
   title={Determining the strange quark mass in Cabibbo-suppressed tau lepton decays},
   volume={533},
   ISSN={0550-3213},
   url={http://dx.doi.org/10.1016/S0550-3213(98)00511-2},
   DOI={10.1016/s0550-3213(98)00511-2},
   number={1–3},
   journal={Nuclear Physics B},
   publisher={Elsevier BV},
   author={Chetyrkin, K.G. and Kühn, J.H. and Pivovarov, A.A.},
   year={1998},
   month=nov, pages={473–493} }

@article{Baikov:2008jh,
    author = "Baikov, P. A. and Chetyrkin, K. G. and Kuhn, Johann H.",
    title = "{Order $\alpha^4(s)$ QCD Corrections to Z and tau Decays}",
    eprint = "0801.1821",
    archivePrefix = "arXiv",
    primaryClass = "hep-ph",
    reportNumber = "SFB-CPP-08-04, TTP08-01",
    doi = "10.1103/PhysRevLett.101.012002",
    journal = "Phys. Rev. Lett.",
    volume = "101",
    pages = "012002",
    year = "2008"
}

@article{PhysRevLett.108.222003,
  title = "{Complete $\mathcal{O}(\alpha_s^4)$ QCD Corrections to Hadronic $Z$ Decays}",
  author = {Baikov, P. A. and Chetyrkin, K. G. and K\"uhn, J. H. and Rittinger, J.},
  journal = {Phys. Rev. Lett.},
  volume = {108},
  issue = {22},
  pages = {222003},
  numpages = {4},
  year = {2012},
  month = May,
  publisher = {American Physical Society},
  doi = {10.1103/PhysRevLett.108.222003},
  url = {https://link.aps.org/doi/10.1103/PhysRevLett.108.222003}
}

@article{Pich_2021,
   title="{Precision physics with inclusive QCD processes}",
   volume={117},
   ISSN={0146-6410},
   url={http://dx.doi.org/10.1016/j.ppnp.2020.103846},
   DOI={10.1016/j.ppnp.2020.103846},
   journal={Progress in Particle and Nuclear Physics},
   publisher={Elsevier BV},
   author={Pich, Antonio},
   year={2021},
   month=mar, pages={103846} }

@article{BRAATEN1992581,
title = "{QCD analysis of the tau hadronic width}",
journal = {Nuclear Physics B},
volume = {373},
number = {3},
pages = {581-612},
year = {1992},
issn = {0550-3213},
doi = {https://doi.org/10.1016/0550-3213(92)90267-F},
url = {https://www.sciencedirect.com/science/article/pii/055032139290267F},
author = {E. Braaten and S. Narison and A. Pich}
}

@article{Surguladze:1990tg,
    author = "Surguladze, Levan R. and Samuel, Mark A.",
    title = "{Total hadronic cross-section in $e^+ e^-$ annihilation at the four loop level of perturbative QCD}",
    reportNumber = "OSU-RN-250",
    doi = "10.1103/PhysRevLett.66.560",
    journal = "Phys. Rev. Lett.",
    volume = "66",
    pages = "560--563",
    year = "1991",
    note = "[Erratum: Phys.Rev.Lett. 66, 2416 (1991)]"
}

@article{Gorishnii:1990vf,
    author = "Gorishnii, S. G. and Kataev, A. L. and Larin, S. A.",
    title = "{The $O(\alpha^{3}_{s})$-corrections to $\sigma_{tot}(e^{+}e^{-}\rightarrow hadrons)$ and $\Gamma(\tau^{-} \rightarrow \nu_{\tau} + hadrons)$ in QCD}",
    reportNumber = "UM-TH-91-01",
    doi = "10.1016/0370-2693(91)90149-K",
    journal = "Phys. Lett. B",
    volume = "259",
    pages = "144--150",
    year = "1991"
}

@article{Becchi:1980vz,
    author = "Becchi, C. and Narison, Stephan and de Rafael, E. and Yndurain, F. J.",
    title = "{Light Quark Masses in Quantum Chromodynamics and Chiral Symmetry Breaking}",
    reportNumber = "CERN-TH-2920",
    doi = "10.1007/BF01546328",
    journal = "Z. Phys. C",
    volume = "8",
    pages = "335",
    year = "1981"
}

@article{Chetyrkin:1979bj,
    author = "Chetyrkin, K. G. and Kataev, A. L. and Tkachov, F. V.",
    title = "{Higher Order Corrections to $\sigma_{tot} (e^+ e^- \rightarrow Hadrons)$ in Quantum Chromodynamics}",
    reportNumber = "IYaI-P-0126",
    doi = "10.1016/0370-2693(79)90596-3",
    journal = "Phys. Lett. B",
    volume = "85",
    pages = "277--279",
    year = "1979"
}

@inbook{king2025righthandedneutrinosseesawmodels,
    author = "King, Stephen F.",
    title = "{Right-handed neutrinos: seesaw models and signatures}",
    eprint = "2502.07877",
    archivePrefix = "arXiv",
    primaryClass = "hep-ph",
    month = "2",
    year = "2025"
}

@article{Bernlochner:2024xiz,
    author = "Bernlochner, Florian U. and Fedele, Marco and Kretz, Tim and Nierste, Ulrich and Prim, Markus T.",
    title = "{Model independent bounds on heavy sterile neutrinos from
                  the angular distribution of $B\to D^{*}\ell \nu$ decays}",
    eprint = "2410.11945",
    archivePrefix = "arXiv",
    primaryClass = "hep-ph",
    reportNumber = "TTP24-041",
    doi = "10.1007/JHEP01(2025)040",
    journal = "JHEP",
    volume = "01",
    pages = "040",
    year = "2025"
}

@article{Asaka_2005,
   title="{The $\nu$MSM, dark matter and baryon asymmetry of the universe}",
   volume={620},
   ISSN={0370-2693},
   url={http://dx.doi.org/10.1016/j.physletb.2005.06.020},
   DOI={10.1016/j.physletb.2005.06.020},
   number={1–2},
   journal={Physics Letters B},
   publisher={Elsevier BV},
   author={Asaka, Takehiko and Shaposhnikov, Mikhail},
   year={2005},
   month=jul, pages={17–26} }

@article{Asaka_2005_2,
   title={The $\nu$MSM, dark matter and neutrino masses},
   volume={631},
   ISSN={0370-2693},
   url={http://dx.doi.org/10.1016/j.physletb.2005.09.070},
   DOI={10.1016/j.physletb.2005.09.070},
   number={4},
   journal={Physics Letters B},
   publisher={Elsevier BV},
   author={Asaka, Takehiko and Blanchet, Steve and Shaposhnikov, Mikhail},
   year={2005},
   month=dec, pages={151–156} }

@article{Yanagida:1979as,
    author = "Yanagida, Tsutomu",
    editor = "Sawada, Osamu and Sugamoto, Akio",
    title = "{Horizontal gauge symmetry and masses of neutrinos}",
    reportNumber = "KEK-79-18-95",
    journal = "Conf. Proc. C",
    volume = "7902131",
    pages = "95--99",
    year = "1979"
}

@article{Fukugita:1986hr,
    author = "Fukugita, M. and Yanagida, T.",
    title = "{Baryogenesis Without Grand Unification}",
    reportNumber = "RIFP-641",
    doi = "10.1016/0370-2693(86)91126-3",
    journal = "Phys. Lett. B",
    volume = "174",
    pages = "45--47",
    year = "1986"
}

@article{Davidson:2002qv,
    author = "Davidson, Sacha and Ibarra, Alejandro",
    title = "{A Lower bound on the right-handed neutrino mass from leptogenesis}",
    eprint = "hep-ph/0202239",
    archivePrefix = "arXiv",
    reportNumber = "OUTP-02-10P, IPPP-02-16, DCPT-02-32",
    doi = "10.1016/S0370-2693(02)01735-5",
    journal = "Phys. Lett. B",
    volume = "535",
    pages = "25--32",
    year = "2002"
}

@article{Belle:2023bwv,
    author = "Prim, M. T. and others",
    collaboration = "Belle",
    title = "{Measurement of differential distributions of $B\rightarrow D^{*} \ell\bar{\nu}_\ell $ and implications on $|V_{cb}|$}",
    eprint = "2301.07529",
    archivePrefix = "arXiv",
    primaryClass = "hep-ex",
    reportNumber = "Belle Preprint 2022-34; KEK Preprint 2022-47",
    doi = "10.1103/PhysRevD.108.012002",
    journal = "Phys. Rev. D",
    volume = "108",
    number = "1",
    pages = "012002",
    year = "2023"
}

@article{Minkowski:1977sc,
    author = "Minkowski, Peter",
    title = "{$\mu \to e\gamma$ at a Rate of One Out of $10^{9}$ Muon Decays?}",
    reportNumber = "Print-77-0182 (BERN)",
    doi = "10.1016/0370-2693(77)90435-X",
    journal = "Phys. Lett. B",
    volume = "67",
    pages = "421--428",
    year = "1977"
}

@article{Robinson_2019,
   title="{Right-handed neutrinos and $R(D^{*})$}",
   volume={2019},
   ISSN={1029-8479},
   url={http://dx.doi.org/10.1007/JHEP02(2019)119},
   DOI={10.1007/jhep02(2019)119},
   number={2},
   journal={Journal of High Energy Physics},
   publisher={Springer Science and Business Media LLC},
   author={Robinson, Dean and Shakya, Bibhushan and Zupan, Jure},
   year={2019},
   month=feb }

@article{Gratrex:2015hna,
    author = "Gratrex, James and Hopfer, Markus and Zwicky, Roman",
    title = "{Generalised helicity formalism, higher moments and the $B \to K_{J_K}(\to K \pi) \bar{\ell}_1 \ell_2$ angular distributions}",
    eprint = "1506.03970",
    archivePrefix = "arXiv",
    primaryClass = "hep-ph",
    reportNumber = "CP3-ORIGINS-2015-017, DIAS-2015-17, CP3-Origins-2015-017 DNRF90, DIAS-2015-17",
    doi = "10.1103/PhysRevD.93.054008",
    journal = "Phys. Rev. D",
    volume = "93",
    number = "5",
    pages = "054008",
    year = "2016"
}

@article{de_Blas_2020,
   title="{HEPfit: a code for the combination of indirect and direct constraints on high energy physics models}",
   volume={80},
   ISSN={1434-6052},
   url={http://dx.doi.org/10.1140/epjc/s10052-020-7904-z},
   DOI={10.1140/epjc/s10052-020-7904-z},
   number={5},
   journal={The European Physical Journal C},
   publisher={Springer Science and Business Media LLC},
   author={de Blas, J. and Chowdhury, D. and Ciuchini, M. and Coutinho, A. M. and Eberhardt, O. and Fedele, M. and Franco, E. and di Cortona, G. Grilli and Miralles, V. and Mishima, S. and Paul, A. and Peñuelas, A. and Pierini, M. and Reina, L. and Silvestrini, L. and Valli, M. and Watanabe, R. and Yokozaki, N.},
   year={2020},
   month=may }

@article{Aoki:2023qpa,
    author = "Aoki, Y. and Colquhoun, B. and Fukaya, H. and Hashimoto, S. and Kaneko, T. and Kellermann, R. and Koponen, J. and Kou, E.",
    collaboration = "JLQCD",
    title = "{$B \rightarrow D^{*}\ell \nu_\ell $ semileptonic form factors from lattice QCD with M{\"o}bius domain-wall quarks}",
    eprint = "2306.05657",
    archivePrefix = "arXiv",
    primaryClass = "hep-lat",
    reportNumber = "KEK-CP-393, OU-HET-1186",
    doi = "10.1103/PhysRevD.109.074503",
    journal = "Phys. Rev. D",
    volume = "109",
    number = "7",
    pages = "074503",
    year = "2024"
}

@article{Harrison:2023dzh,
    author = "Harrison, Judd and Davies, Christine T. H.",
    collaboration = "HPQCD, (HPQCD Collaboration){\textdaggerdbl}",
    title = "{$B\rightarrow D^{*}$ and $B_s \rightarrow D_s^{*}$ vector, axial-vector and tensor form factors for the full $q^2$ range from lattice QCD}",
    eprint = "2304.03137",
    archivePrefix = "arXiv",
    primaryClass = "hep-lat",
    doi = "10.1103/PhysRevD.109.094515",
    journal = "Phys. Rev. D",
    volume = "109",
    number = "9",
    pages = "094515",
    year = "2024"
}

@article{FermilabLattice:2021cdg,
    author = "Bazavov, A. and others",
    collaboration = "Fermilab Lattice, MILC",
    title = "{Semileptonic form factors for $B\rightarrow D^*\ell \nu $ at nonzero recoil from $2+1$-flavor lattice QCD: Fermilab Lattice~and~MILC~Collaborations}",
    eprint = "2105.14019",
    archivePrefix = "arXiv",
    primaryClass = "hep-lat",
    reportNumber = "FERMILAB-PUB-21-261-T~, FERMILAB-PUB-21/261-T",
    doi = "10.1140/epjc/s10052-022-10984-9",
    journal = "Eur. Phys. J. C",
    volume = "82",
    number = "12",
    pages = "1141",
    year = "2022",
    note = "[Erratum: Eur.Phys.J.C 83, 21 (2023)]"
}

@article{PhysRevLett.43.668,
  title = "{Higher-Order Quantum Chromodynamic Corrections in ${e}^{+}{e}^{\ensuremath{-}}$ Annihilation}",
  author = {Dine, Michael and Sapirstein, Jonathan},
  journal = {Phys. Rev. Lett.},
  volume = {43},
  issue = {10},
  pages = {668--671},
  numpages = {0},
  year = {1979},
  month = {Sep},
  publisher = {American Physical Society},
  doi = {10.1103/PhysRevLett.43.668},
  url = {https://link.aps.org/doi/10.1103/PhysRevLett.43.668}
}

@article{PhysRevLett.44.560,
  title = "{Analytic Calculation of Higher-Order Quantum-Chromodynamic Corrections in ${e}^{+}{e}^{\ensuremath{-}}$ Annihilation}",
  author = {Celmaster, William and Gonsalves, Richard J.},
  journal = {Phys. Rev. Lett.},
  volume = {44},
  issue = {9},
  pages = {560--564},
  numpages = {0},
  year = {1980},
  month = {Mar},
  publisher = {American Physical Society},
  doi = {10.1103/PhysRevLett.44.560},
  url = {https://link.aps.org/doi/10.1103/PhysRevLett.44.560}
}

@article{Gorishnii:1988bc,
    author = "Gorishnii, S. G. and Kataev, A. L. and Larin, S. A.",
    title = "{Next-To-Leading $\mathcal{O}(\alpha_s^3)$ QCD Correction to $\sigma_{tot} (e^+ e^- \rightarrow Hadrons)$: Analytical Calculation and Estimation of the Parameter Lambda (MS)}",
    reportNumber = "JINR-E2-88-254",
    doi = "10.1016/0370-2693(88)90532-1",
    journal = "Phys. Lett. B",
    volume = "212",
    pages = "238--244",
    year = "1988"
}

@article{Kataev:1990kv,
    author = "Kataev, A. L.",
    editor = "Narison, Stephan",
    title = "{Next next-to-leading perturbative QCD corrections: The Current status of investigations}",
    reportNumber = "PM-90-41",
    doi = "10.1016/0920-5632(91)90669-6",
    journal = "Nucl. Phys. B Proc. Suppl.",
    volume = "23",
    pages = "72--90",
    year = "1991"
}

@article{kretz:wpaper,
    author = "Kretz, Tim and Nierste, Ulrich",
    title = "{QCD corrections to charged-current decays with Heavy Sterile Neutrinos in initial or final state and their impact on $\tau$ decays}",
    eprint = "2512.00476",
    archivePrefix = "arXiv",
    primaryClass = "hep-ph",
    reportNumber = "P3H--25--103, TTP25-049",
    month = "11",
    year = "2025"
}

\end{document}